\documentstyle [12 pt] {article}

\begin{document}
\twocolumn

\Huge
{\bf Observatories}
\normalsize


\section {Introduction} 

An astronomical observatory is a building, installation, 
or institution dedicated to the systematic and regular
observation of celestial objects for the purpose of 
understanding their physical nature, or for purposes of time 
reckoning and keeping the calendar.  The work at a {\em bona 
fide} observatory constitutes a main activity, not just an 
incidental one.  Observatories gather data on celestial 
positions, motions, luminosities, or chemical composition 
which can be used as a basis for theories that describe {\em 
what} these celestial bodies are, where they are to be 
located in the sky, or how they evolve.

While the ancient Egyptians, Babylonians, Chinese, neolithic Britons,
Greeks and Mayans made many observations of scientific and historical
interest, their efforts were not carried out at observatories {\em per
se}.  In this article we shall concentrate on installations that were
intended to be {\em bona fide} research centers. 

\section {Observatories in Islam}

There was a strong motivation in medieval Islam to 
preserve the knowledge of the Greeks and to add to it, if 
possible.  In a large city the existence of a hospital, a 
university, and a public library was considered sensible and 
desirable.  Within this context, wherein knowledge and piety 
were highly encouraged and intertwined, the tradition of 
observatory building began during the reign of al-Mam\={u}n, the 
seventh Abbasid caliph (813-833 AD). He organized a 
scientific academy in Baghdad called the House of Wisdom, 
which included a library and observatory.  This was the most 
significant scientific endeavor since the establishment of 
the Alexandrian Museum in the 3rd century BC.  A second 
observatory was built on the plains of Tadmor in Syria.  
al-Mam\={u}n's astronomers measured the obliquity of the ecliptic 
(obtaining a value of 23$\rm ^o$ 33$^{\prime}$), concluded that the 
precession of the equinoxes amounted to 54 arcsec per year, 
determined that the Earth was 6500 miles in diameter, and 
produced tables of motion of the planets based on Ptolemy's 
work, but updated on the basis of new observations.  
Astronomical work continued in Baghdad under the ninth 
Abassid caliph, but then ceased.

A second scientific academy, the Hall of Wisdom, was 
established in Cairo in 1005.  It lasted until 1171. An 
observatory was begun there in 1120 and nearly completed, but was 
ordered destroyed by the caliph in 1125 after the death of 
the vizier in charge of the observatory and its instruments.  
Apparently, serious conflicts had arisen between this 
vizier and the caliph, and it did not help that the 
superstitious populace presumed that trying to understand the 
motions of the planets was akin to delving into magic.

The Istanbul Observatory, completed in 1577, met a 
similar demise.  Political back stabbing and superstitions 
regarding astronomy led to that observatory's destruction by 
1580.

The two most successful and extensive Islamic 
observatories were those at Mar\={a}gha (in Azerbaijan, in 
modern-day Iran) and at Samarkand, in modern-day Uzbekistan.  
The Mar\={a}gha Observatory was built under the direction of 
Na\d{s}\={\i}r al D\={\i}n al-\d{T}\={u}s\={\i} (1201-1274).  
It was situated on the 
flattened top of a hill, covering 400 by 150 meters.  
Completed in 1264, it was active at least until 1304, perhaps 
until 1316.  It was the first observatory whose activity did 
not greatly diminish after the death of its founder.  While 
the motivation for the construction of the observatory was 
{\em astrological}, the prediction of future celestial events 
needed to be carried out on the basis of exact physical 
measurements.  A significant number of instruments was built, 
amongst them equinoctial and solsticial armillary spheres, 
and a mural quadrant of radius 4.3 meters.  The observatory's 
library at one time contained 400,000 volumes.  The principal 
accomplishment at Mar\={a}gha was the compilation of the Ilkhanic 
Tables, which were tables of motion of the Moon, Sun, and 
planets.  These explained some of the major shortcomings of 
Ptolemaic astronomy without, however, suggesting 
heliocentrism.

The most important astronomer of the 15th century was 
Ulugh Beg (1394-1449).  For most of Ulugh Beg's life his 
father was the ruler of Transoxiana, a region situated 
between the River Oxus (Amur Darya) and the River Jaxartes 
(Syr Darya).  This provided the son with the opportunity to 
indulge his passion for astronomy.  As a young man Ulugh Beg 
visited the remains of the Mar\={a}gha Observatory, and he 
endeavored to build an even greater institution in Samarkand.  
As many as 70 astronomers were active there between 1408 and 
1437.  The largest instrument constructed was a sextant of 
radius 40 {\em meters}, a fixed instrument mounted on the 
north-south meridian which could achieve a resolution of 
several arc seconds, a value not to be exceeded until the 
invention of the telescope and the micrometer in the 17th 
century.  This sextant was primarily used for observations of 
the Sun.

On the basis of {\em new} observations, Ulugh Beg and 
his fellow astronomers compiled a catalogue of the brightest 
1000 stars visible at the latitude of Samarkand.  No one 
since Ptolemy (ca. 150 AD) or perhaps as far back as 
Hipparchus had made such extensive stellar observations.  
Ulugh Beg's catalogue was most likely based on observations 
with a zodiacal armillary sphere graduated to 15 arc minutes, 
with interpolation to two-tenths of a mark, or 3 arc minutes.  
The typical positional accuracy in the catalogue is $\pm$ 16 
arc minutes.  

Briefly stated, the purpose of observatories in Islam  
was the production of astronomical ephemerides of the Sun, 
Moon, and planets, and possibly also the compilation of a 
star catalogue.  Royal patronage was required.  The Mar\={a}gha 
and Samarkand Observatories were intended to be permanent 
institutions.  Centuries ago it was realized that 
astronomical research is necessarily open-ended.  Better 
positional accuracy is always desirable, so new instruments 
have to be designed in order for positional accuracy to be 
improved.  Instruments made of metal were found to last 
longer and give more consistent results than instruments made 
of stone or wood.  In order to be useful, observations had to 
be carried out for specific purposes and according to 
specific methods.

\section {China and India}

At the very end of the 17th century one Louis Lecomte 
published his {\em Memoirs and Observations of a Journey 
through China}.  He noted in some detail the vigilant 
activity carried out by astronomers at the Imperial 
Observatory in Beijing: ``Five mathematicians spend every 
night on the tower in watching what passes overhead; one is 
gazing towards the zenith, another to the East, a third to 
the west, the fourth turns his eyes southwards, and a fifth 
northwards, that nothing of what happens in the four corners 
of the world may escape their diligent observation.'' Such 
activity had been going on for 3000 years.  As a result, the 
Chinese had accumulated records pertaining to many hundreds 
of lunar and solar eclipses, observations of novae, 
supernovae, comets, meteor showers, aurorae, naked eye 
sunspots, and even a possible observation of Jupiter's moon 
Ganymede in 364 BC, nearly two millenia before the invention 
of the telescope!  The first clock drive was built by the 
Chinese in 132 AD; it was powered by a constant pressure-head 
of water in a {\em clepsydra}, or water clock.  The Chinese 
astronomer Y\"{u} Xi independently discovered the precession 
of the equinoxes in 320 AD, finding a value of about 72 arc 
seconds per year, two times the value found by Hipparchus.  
Sunspot records in the Chinese annals demonstrate the 
11-year solar cycle.

Just as spices were transported along the caravan 
routes, so too was astronomical knowledge.  Astronomers from 
Persia arrived in China in 719 AD.  The astronomical ties 
between the mid-East and far East became quite extensive 
during the Y\"{u}an Dynasty (1271-1368).  In 1267, only eight 
years after the founding of the Mar\={a}gha observatory, 
blueprints for seven instruments were sent to China.  Actual 
models may have followed.

In 1270 the astronomer Guo Shoujing (1231-1316) built 
the first equatorially mounted instrument.  After the Beijing 
Observatory was re-equipped in 1276 to 1279 under his 
direction, it was equal in stature to the Mar\={a}gha 
Observatory.  The Chinese also operated an observatory at 
Nanjing.

  The Chinese did not appreciate or understand the Greek 
geometrical models of planetary motions or the Arabic use of 
geometry, particularly spherical trigonometry and 
stereographic projection. However, three hundred years 
before Tycho Brahe became convinced that right ascension and
declination were the ``coordinates of the future'', the 
Chinese were fully committed to their use.  Just as the 
influence of the Emperor radiated in all directions, so
the hour circles radiated from the pole, ``like the spars of
an umbrella.'' The Chinese laid out a system of 28 lunar
mansions ({\em xiu}), which were defined by the points at
which these hour circles intersected the celestial equator.

One of the most important events in the history of 
Chinese science was the arrival of the Jesuits in 1600, 
towards the end of the Ming Dynasty (1368-1644).  They had a 
very specific motivation.  By demonstrating the superiority 
of Western science, they hoped to convince the Chinese of the 
superiority of Western religion.  Western science clearly won 
out over the traditional Chinese and Muslim methods of 
predicting the solar eclipses of 15 December 1610 and 21 June 
1629.  In November of 1629 a new calendar bureau was 
established under the direction of the Chinese Christian 
convert Xu Guangqi, who supervised the work of fifty 
astronomers, many of whom converted to Christianity.  
However, in 1664, at the start of the Qing Dynasty 
(1644-1911), politics led to the dismissal of the Jesuits and most 
of the Chinese converts from the calendar bureau.  Some of 
these astronomers were later beheaded.

The Jesuits regained favor in Chinese astronomy in 1669.  
From 1673 to 1676 the Beijing Observatory was re-equipped 
with a new set of Tychonic instruments, which are still in 
place today.  However, the astronomical activities in China
were hindered by several factors: 1) the interests of the
Chinese were limited to calendrical revision; 2) for Catholics,
such as the Jesuits, the Church's prohibition against discussing
heliocentrism was in effect from 1616 to 1757; 3) the popularity
of Tycho's hybrid geo-heliocentric model of the solar system
in the minds of the Jesuits further eclipsed interest in
Copernicanism; and 4) since the most fundamental motivation
of the Jesuits was religious conversion, not scientific progress,
they were not really interested in telescopes and telescopic
observations.  Meanwhile, back in Europe astronomers such as
Jean Dominique Cassini and John Flamsteed (see \S0.5.1 below) were
breaking new ground in the realms of planetary and positional astronomy.

The most significant observatory construction in India was
carried out under the direction of Jai Singh (1686-1743),
a Hindu prince in the court of a Muslim Mogul emperor.
Large instruments of masonry were constructed at Delhi,
Jaipur, Ujjain, Benares, and Mathura.  The largest instrument 
was a sundial 27 meters tall.  Jai Singh wished to follow
in the footsteps of Ulugh Beg.  He hoped to minimize observational
errors by using the largest instruments possible and felt
that portable brass instruments were not the instruments
of choice.  Jai Singh's star catalogue was an update
of Ulugh Beg's star catalogue, but apparently with no
new observations.  4$\rm ^o 8^{\prime}$ were simply added to 
Ulugh Beg's ecliptic longitudes to account for precession 
over 288 years.  
 
\section {Early European observatories}

Astronomical observations were carried out in Moslem 
Spain at Cordova in the 10th century, at Toledo in the 11th 
century, and at Castille in the 13th century under the 
patronage of the Christian King Alfonso X.  The Toledan 
astronomical tables were translated into Latin in the 12th 
century; Copernicus owned a copy of them.  The Alfonsine 
Tables, which form the first state-sponsored astronomical 
ephemerides published for general use, were reprinted as 
late as 1641. 

The first European observatory worthy of the name was 
built by Bernard Walther (1430-1504), a wealthy private 
citizen of Nuremberg, who was at the same time the pupil and 
patron of Regiomontanus (1436-1476).  Together they found 
that the positions of the planets differed to a significant 
degree from the predictions of the Alfonsine Tables.  Their 
observations of the comet of 1472 provided a basis for the 
modern study of comets.  They printed astronomical treatises 
on a printing press set up in Walther's house.  One such 
treatise by Regiomontanus laid out the ``method of lunar 
distances'' for determining longitude at sea.  A significant 
innovation at Walther's observatory was determination of the 
time of observations by mechanical clocks instead of by 
astrolabes or armillary spheres.  Observations with 
Walther's instruments exhibited an improved accuracy, to 
$\pm$ 10 arcmin.

Wilhelm IV (1532-1592), the Landgrave of Hesse-Cassel, 
began systematic observations in 1561.  In 1567 his father, 
the Landgrave of Hesse, died, and the landgraviate was 
divided up amongst the four sons.  Wilhelm was compelled to 
take over the administration of his province, and had to lay 
aside his astronomical endeavors.  In 1575, however, he was 
visited by Tycho Brahe and was inspired to resume observing.

Wilhelm constructed a number of metal instruments, 
amongst them an azimuthal quadrant of radius 0.4 m, a 
torquetrum, a sextant of radius 1.3 m, a quadrant of 1.5 m 
radius, and armillary spheres (see article on {\em 
Pre-~telescopic instrumentation}).  He demonstrated the 
superiority of metal instruments over wooden ones.  Wilhelm's 
most significant innovation was the rotating dome; he 
observed from one built in a tower of his castle.  His star 
catalogue of about 400 stars, though far short of the 
intended number of stars he had hoped to measure, was one of 
the first made in Europe. 

The most significant observatory prior to the invention 
of the telescope was that of Tycho Brahe (1546-1601).  It was 
situated on the 2000 acre island of Hven, in the Danish sound 
between Copenhagen and Elsinore.  There Tycho built 
Uraniborg, the ``castle of the heavens''.  It sat in the 
middle of a square 300 feet on a side and enclosed by a wall 
22 feet high.  The castle itself had two main levels and was 
taller than a seven story building.  In addition to the 
observing rooms and verandahs on the upper level, it 
contained a dozen or so bedrooms, a dining room, library, 
chemical laboratory, and even a jail in the basement.  Begun 
in 1576, it was completed in 1580.  Over the north portal was 
an inscription, carved in stone: {\em Nec fasces nec opes 
sola artis sceptra perennant} (neither wealth nor power, 
but only knowledge, alone, endures).  Uraniborg was the first 
scientific research institute in Renaissance Europe, and 
Tycho was the first full-time scientist.

At Uraniborg Tycho and his assistants used a dozen 
instruments.  The one that produced the most accurate stellar 
positions was a mural quadrant of radius 2 m, which was 
located in the southwest room on the ground floor.  A mural 
quadrant, of course, is attached to a wall, fixed on the 
celestial meridian.  Being a quadrant located at latitude 
55$\rm ^o 54^{\prime} 26^{\prime \prime}$, 
it could not be used to observe stars north 
of $\delta$ = 55.9 degrees.  Its fine construction and the 
use of transversals\footnote {Transversal dot lines in 
graduations were first introduced by Richard Cantzlar in 
1552 or 1553.  This was a variant of the zigzag line system 
introduced by Johann Hommel (1518-1562), from whom Tycho 
Brahe obtained it.  The adoption of concentric circles ruled 
with different integral numbers of lines was suggested by P. 
Nu$\rm \tilde{n}$ez in 1542, which led to P. Vernier's 
invention in 1631.  The micrometer and filar micrometer
were invented by William Gascoigne about 1640, though they
only came into more general use in England, France, and
Holland in the 1670's.} allowed the determination of 
declinations with a resolution of 10 arc sec.   Perhaps 
because of the success of this very instrument, a transition 
began at this time from positional instruments which could be 
used near the celestial meridian to those which were 
explicitly designed to do so. 

An auxiliary observatory, Stjerneborg,  was built in 
1584 one hundred feet to the south of the main building.  There were 
five new instruments, the largest of which was an equatorial 
armillary made of iron, which was 2.9 m in diameter and could 
be used to measure declinations with a resolution of 15 arc 
sec.  Each of these five instruments was situated in a 
subterranean crypt.  Three were sheltered by folding roofs. 
Two had revolving domes.

From observations at Hven Tycho discovered four inequalities in the Moon's
motion, two in longitude (the ``variation'' and the annual equation with a
period of one solar year), and two in latitude.  His observations of Mars
were used later by Kepler to discover the elliptical nature of planetary
orbits.  From observations of 1588 to 1591, he produced a catalogue of 777
stars with an improvement in positional accuracy of an order of magnitude. 
More than one of Tycho's instruments produced positional accuracies
for his nine principal reference stars of $\pm$ 0.6 arc min.
The average uncertainty of the
(other) brighter stars in Tycho's catalogue is 1.9 arc min in ecliptic
longitude and 1.2 arc min in ecliptic latitude.  For the fainter stars
the uncertainties amount to 2.8 and 2.6 arc min, respectively.

Tycho was able to afford such a magnificent observatory 
because his annual income from various fiefdoms was the 
equivalent of one percent of the Danish government's income.  
This continued for a period of nearly 30 years.  Such funding 
in today's currency would support an observatory more 
expensive than the Hubble Space Telescope and ESO's Very 
Large Telescope combined.  It was not just money, of course, 
that led to the improvement in instrumentation.  More than 
one of Tycho's metal instruments required the workmanship of 
five or six people and three years.

Tycho's fortunes, both literally and figuratively, began 
to decline in 1588 after the death of Tycho's patron, the 
Danish king Frederick II.  Tycho left Hven in 1597 and his
observatory rapidly fell into decline.

Johannes Hevelius (1611-1687), the son of a prosperous brewer of Danzig
(Gdansk), built what was for a short time the world's leading observatory. 
He copied a number of Tycho's instruments, constructing many quadrants and
sextants of wood or copper.  For observing the moon and planets he
constructed refractors up to 150 feet in focal length, in which the
objective was mounted on a tall mast.  However, for stellar positions he
clearly preferred the unaided eye, as he explicitly stated in his {\em
Machina Coelestis} (1673). This precipitated a controversy with Robert
Hooke, who was a strong advocate of telescopic sights.  Edmund Halley was
compelled to visit Danzig in 1679 to try to resolve the controversy.  For
two months Halley observed with a telescope fitted with sights while
Hevelius used naked eye instruments. Their observations were of equal
accuracy, slightly less than one arc minute.  The {\em Selenographia}
(1647) of Hevelius was a milestone in lunar mapping.  His posthumous star
catalogue (1690) contains the positions of 1564 stars.

\section {The rise of national observatories}

\subsection{First era: 1576-1725}
We may distinguish three eras of construction of 
national observatories.  Tycho's island observatory was, in 
effect, a national observatory, but the Paris Observatory 
(established 1667) was the prototype of the national 
observatories that followed.  Astronomers in Paris, along 
with those at Royal Greenwich Observatory (RGO, 1675), Berlin 
(1701), and St. Petersburg (1725), were dedicated to 
practical matters of national importance: improving 
navigation (especially the determination longitude at sea), 
geodesy, calendar reform,  producing accurate stellar 
coordinates, and the determination of ephemerides of the Sun, 
Moon, and planets.  The establishment of scientific academies 
such as the {\em Accademia dei Lincei} (Rome, 1603), the Royal 
Society (London, 1662), and the {\em Acad\'{e}mie des Sciences} 
(Paris, 1666) testified to the importance and prestige of 
scientific work.

The Paris and Greenwich observatories stand out as institutional
models upon which many subsequent observatories were based.
Paris was a reflection of the splendor of the court of Louis XIV,
and many significant astronomers were associated with it.  Jean
Dominique Cassini (1625-1712), his son, grandson, and great-grandson,
all of whom were unofficial directors at Paris, comprised a most
notable astronomy dynasty.  The first Cassini discovered four moons
of Saturn and the division of the ring system named after him.
Paris astronomers used extremely long focal length refractors (up to
136 feet), mounted on tall masts, to observe the planets and Moon.
Ole R\"{o}mer demonstrated the finite nature of the speed of light
from observations of Jupiter's Galilean moons.  Paris astronomers
carried out the first geodetic stuveys, covering the full arc of
meridian of France by 1700.  They later sent expeditions to Peru
and Lappland.  After Nicolas de Lacaille went to South Africa in 1751
to measure the positions of bright stars and to do further geodetic
work, it was proven that the Earth's figure was oblate.

From the outset both Paris and Greenwich were mandated to aid navigation
through the production of accurate star catalogs, ephemerides of the Sun,
Moon, and planets, and the production of nautical almanacs.  The French
{\em Connaissance de Temps} first appeared in 1679.  In 1795 its 
production
became the responsibility of the newly created {\em Bureau des
Longitudes}, which supervised the work of the Paris Observatory.  In
England the Board of Longitude supervised the work of the RGO from 1714 to
1828.  By then, with the existence of accurate star positions,
ephemerides, and the invention of the marine chronometer, the problem of
determining one's geographical position at sea was essentially solved. 

The buildings, staff size, and funding level of the RGO were, until the
twentieth century, modest compared to Paris.  The Astronomers Royal
Flamsteed, Halley, Bradley, and especially George Biddell Airy (1801-1892)
were amongst the most important astronomers of their day.  Airy organized
the observatory as an astronomical factory with himself as factory
director and sought to carry out the observatory's mission (data
acquisition, reduction, and publication of results) as accurately and
efficiently as possible.  Today's service observing and remote observing
operations are direct descendants of such efforts. 

\subsection {Second era: 1820-1918}
The second era of national observatory building was 
characterized by offshoots from previous national 
observatories (Royal Observatory Cape, South Africa, 1820), 
by newer observatories of younger nations (United States 
Naval Observatory, 1839), and the rise of astrophysical 
observatories (Potsdam, Prussia, 1874).  Other national 
observatories of the second era include Pulkovo (Russia, 
1839), the Chilean National Observatory (1852), the Argentine 
National Observatory (1870), the Smithsonian Astrophysical 
Observatory (USA, 1891), and Canada's Dominion Observatory 
(1903) and Dominion Astrophysical Observatory (1918).

Of~nineteenth~century~observatories, Pulkovo deserves 
special mention.  Just as Tycho was given {\em carte blanche} 
by his sovereign to build the finest observatory in the 
world, so too was Wilhelm Struve (1793-1864) amply funded by 
Tsar Nicholas I.  Struve ordered an Ertel transit instrument 
(for the determination of absolute right ascensions), an 
Ertel vertical circle (for the determination of absolute 
declinations), a Repsold meridian circle (for the 
determination of differential stellar coordinates) and prime 
vertical transit (for the determination of aberration and 
nutation), a Merz \& Mahler heliometer (for the determination 
of angular sizes or angular separations), and the 15-inch 
Merz \& Mahler refractor (for the discovery and measurement of 
double stars).  This refractor was the largest telescope in 
the world at that time.  A twin was built for Harvard College 
Observatory in 1847.

The instruments listed above testify to the maturation 
of positional astronomy as a field of endeavor.   Not only 
did Pulkovo astronomers measure some of the first stellar 
parallaxes and discover many double stars, they also produced 
the most accurate values for the constants of precession, 
aberration, and nutation -- values not superceded until 1964.

 It is also important to mention that progress in 
astronomy depended very greatly on progress in instrument 
making and optics.  Thomas Tompion, George Graham, John Bird, 
Jesse Ramsden, Joseph Fraunhofer, Edward Troughton, Carl 
Zeiss, Alvan Clark, Alvan Graham Clark, Howard Grubb,
and George Ritchey are some of the men who made 
significant contributions during this era.

\subsection {Third era: 1956-present}
The present era of national observatory building is 
characterized by national or international consortia, large 
budgets, and the investigation of celestial objects at all 
electromagnetic wavelengths.  From ground-based observatories 
we may investigate optical, infrared, radio, and 
submillimeter waves.  Because of absorption by water vapor in 
the Earth's atmosphere of certain infrared and submillimeter 
waves, we must make some observations from balloons and 
aircraft.  Ultraviolet, X-ray, and gamma-ray astronomy {\em 
must} be carried out from satellites.

Significant national observatories of the third era 
include the National Radio Astronomy Observatory (USA, 1956), 
Kitt Peak National Observatory (USA, 1957), NRAO (Australia, 
1959), Cerro-Tololo Inter-American Observatory (Chile, 1963), 
European Southern Observatory (Chile, 1964), Anglo-Australian 
Observatory (1967), the Kuiper Airborne Observatory 
(USA, 1975-1995), and the Space Telescope Science Institute (1981).  

\section {Private observatories}

Not all significant astronomical work is carried out at 
observatories funded by federal governments, universities, 
and research foundations.  In the late eighteenth century and
throughout the nineteenth century various 
(usually wealthy) individuals established their own private 
observatories.

The most accomplished private astronomer of all time was the
German-English astronomer William Herschel (1738-1822).  He
pioneered the production of larger and larger {\em reflecting}
telescopes (up to 48 inches in diameter), whose light gathering
power enabled him to establish the field of galactic astronomy.
He carried out deeper and deeper surveys of the northern sky,
discovered hundreds of binary stars and many hundreds of nebulae.
His ``star gauges'' showed that our solar system is situated
in a large flattened disk of stars.  From proper motion data
he showed that our solar system is moving toward the constellation
Hercules.  He proved that gravity works in the stellar realms by
demonstrating that members of a binary star pair can be observed
to revolve about their center of mass.  He laid the groundwork for
studies of the evolution of stars and stellar systems.

After Herschel's discovery of the planet Uranus in 1781 he was able
to give up his first profession (music) and moved to Slough (near
Windsor), where his only occasional duty was to give private star
parties for members of Royalty.  Herschel's son John followed in
his father's footsteps and took an 18-inch, 20 foot focal length,
reflector to South Africa in the 1830's, where he conducted a
southern sky survey, all done by eye and telescope, of course.
The nebulae discovered by the Herschels formed the basis of 
Dreyer's {\em New General Catalogue} (1888).

Following Herschel's discovery of Uranus, the
German astronomer Johann Hieronymus Schr\"{o}ter (1745-1816)  quit his
post in Hanover and moved to the small town of Lilienthal to devote
himself to observational astronomy.  He built what was then the largest
observatory on the European continent, and was assisted by Karl Ludwig
Harding (who discovered the third asteroid Juno in 1804) and Friedrich
Wilhelm Bessel.  Schr\"{o}ter was also closely associated with H. W. M.
Olbers.  Schr\"{o}ter made detailed drawings of features on the lunar
surface and believed he detected changes in the structure of some of those
features. He inspired many observers to dedicate themselves to planetary
studies.

William Parsons, the Third Earl of Rosse (1800-1867) 
built a six foot diameter reflecting telescope in 
Parsonstown, Ireland, which revealed the spiral nature of 
some nebulae (i.e. galaxies).

William Huggins (1824-1910) and his wife Margaret 
Lindsay Murray established a private observatory at Tulse 
Hill, near London, where they observed the spectra of stars 
and other celestial objects.  They proved that some nebulae 
are star clusters, and were the first to prove that others 
are composed entirely of glowing gas.

In 1868, simultaneously with the Frenchman Jules Janssen,
the English amateur Norman Lockyer (1836-1920) 
used a spectroscope attached to a refractor to observe solar 
prominences outside of a total solar eclipse.  He was 
subsequently appointed head of the Solar Physics Observatory 
at South Kensington.  When he retired in 1911 he established 
his own observatory at Salcombe Regis.

The American Henry Draper (1837-1882) was a pioneer of 
astrophotography.  His widow funded significant spectroscopic 
research carried out by astronomers at the Harvard College 
Observatory.

George Ellery Hale (1868-1938), the son of a wealthy 
Chicago industrialist, invented the spectroheliograph while 
an undergraduate at the Massachusetts Institute of 
Technology.  He established a private solar observatory in 
his parents' backyard in Kenwood, then a suburb of Chicago.  
This later became part of the University of Chicago.  Hale 
became the greatest observatory entrepreneur of all time, 
establishing the Yerkes Observatory (1897) and the Mt. Wilson 
Observatory (1904).  He was the driving force behind the 
establishment of Palomar Observatory (1948).

At the end of the twentieth century, due to advances in 
electronics and imaging devices, many amateurs obtain 
significant hard data for professional research projects.  
Photoelectric and CCD photometry by amateurs has been very 
important for variable star research and has even led to the 
discovery of a new class of pulsating stars.  Astrometry from 
CCD imagery routinely provides data on recently discovered 
asteroids, which allows the calculation of their orbits.  
Until very recently the most successful discoverer of 
supernovae in other galaxies was the Australian amateur 
Robert Evans.

\section {Mountaintop observatories and the modern era}

As a result of his optical researches, Isaac Newton 
realized that the Earth's atmosphere acts as a lens of sorts 
and that one could achieve much better astronomical image 
clarity by siting one's telescope on ``some lofty mountain''.

In 1741 the French astronomer Fran\c{c}ois de Plantade 
carried out the first mountaintop astronomical observations
at Sencours Pass, at the foot of 
Pic du Midi in the French Pyrenees, at an altitude of 2300 
meters.  Unfortunately, he succumbed to altitude sickness and 
died.

The next small step leading to improved observing was 
taken with the establishment in 1783 of the Dunsink 
Observatory, an institute of Trinity College, Dublin.  The 
observatory was intentionally sited outside the city (8 km).  
Also, it had the first fully functional rotating 
hemispherical dome, and the primary telescope was mounted on 
a pier which was structurally (and therefore vibrationally) 
isolated from the walls of the building.

In 1856 the Scottish astronomer Charles Piazzi Smyth 
spent nearly four months at Tenerife in the Canary Islands 
and demonstrated that mountaintop observing was both 
desirable (for image clarity) and viable (for humans).

The first mountaintop observatory intended to be 
permanent was Lick Observatory, which was completed in 1888.  
Situated at the 1280 m summit of Mt. Hamilton, near San Jose, 
California, it boasted a 36-inch refractor by Alvan Clark, 
then the world's largest.  After the spectacular success of 
the Mt. Wilson Observatory and the Palomar Observatory, it 
became clear that the best seeing was obtained at mountaintop 
sites within 50 miles of the ocean.  Because of the proximity 
of the ocean, one gets a smooth, laminar flow of air over the 
coastal mountains when the prevailing wind conditions are in 
effect.  This has been borne out by the establishment of many 
other observatories during the second half of the twentieth 
century.
 
The highest major observatory in the world (at 4205 m) 
is at Mauna Kea on the island of Hawaii.  From a very modest 
start in 1964, Mauna Kea's astronomical installations 
expanded greatly and now include the University of Hawaii 
2.2-m telescope, the Canada-France-Hawaii Telescope (3.6-m), 
the NASA Infrared Telescope Facility (3.0-m), the United 
Kingdom Infrared Telescope (3.8-m), the James Clerk Maxwell 
submillimeter telescope (15-m), the Caltech Submillimeter 
Observatory (10.4-m), one of the nine elements of the Very 
Long Baseline Array (25-m radio telescope), the two Keck 
telescopes (each 10-m), the Japanese optical and infrared 
telescope Subaru (8.3-m), the northern Gemini telescope (8.1-
m), and the Smithsonian Astrophysical Observatory 
Submillimeter Array.

Presently, the highest established telescopes are the 
twin 0.7-m reflectors of the Meyer-Womble Observatory at Mt. Evans, 
Colorado (elevation 4313 m).  A still higher observatory is 
being built at 5000 m about 40 km east of San Pedro de 
Atacama in Chile.  This is the centimeter~-wavelength
Cosmic Background Imager, 
which will investigate the small ripples in the 3 K 
background radiation leftover from the Big Bang.  This observatory
will have an oxygen-enhanced environment for the operators.

At the time of the publication of this article there 
will be more than 90 optical or infrared telescopes in 
operation around the world with diameters of 1.5 meters or 
larger (see figure).  This includes the now single-mirror 
6.5-m MMT at Mt. Hopkins, Arizona, the twin 6.5-m reflectors 
of the Magellan Project (situated at Las Campanas, Chile), 
the twin 8.1-m Gemini telescopes (one at Mauna Kea, Hawaii, 
the other at Cerro Pachon, Chile), the four 8.2-m elements of 
the European Southern Observatory's Very Large Telescope (at 
Cerro Paranal, Chile), Japan's 8.3-m Subaru Telescope (at 
Mauna Kea), the two 8.4-m elements of the Large Binocular 
Telescope (at Mt. Graham, Arizona), the two 10-m Keck 
telescopes at Mauna Kea, and the 11-m Hobby-Eberly 
spectroscopic telescope at Mt. Locke, Texas.  In fact, by the 
turn of the 21st century more than half the {\em area} of 
telescopes larger than 1.0 meters will be in telescopes 
larger than 8 meters.

Given the expense of such facilities (measured in units 
of hundreds of millions of dollars), one might naturally ask 
why so many large telescopes are being built. It is a 
reflection of the blossoming of extragalactic observational 
astronomy, the desire to give hard data to cosmologists for 
testing their models, and the desire to discover and study 
intrinsically faint objects such as asteroids in the outer 
solar system, brown dwarf stars,  and extrasolar planets.

The design and construction of these state-of-the-art 
telescopes has required the development of computer-
controlled mirror mounts, better ventilated domes (to 
eliminate as much as possible the degradation of the seeing 
due to air in the dome), thin mirrors (as in the case of most 
of the telescopes mentioned above) or multi-mirror systems 
(36 hexagonal segments in the case of each of the Keck 
telescopes, 91 segments in the case of the Hobby-Eberly 
Telescope).  Modern instrumentation often uses techniques 
borrowed from spy satellites to achieve diffraction-limited 
performance.  Whereas photographic plates a century ago had a 
quantum efficiency (QE) of less than one percent and 
photomultiplier tubes typically had peak QE's of a few 
percent, modern instrumentation can achieve QE's of 70 
percent.  Such instruments, fed by the large objectives of 
the present generation of telescopes, are producing many 
spectacular results.  Furthermore, modern computing 
capabilities are allowing digital celestial surveys which 
would have been technically impossible only a few years ago.  

\vspace {5 mm}

\parindent = 0mm

Dick S J 1990 Pulkovo Observatory and the national 
observatory movement: an historical overview In: Lieske, J H, 
Abalakin, V K (eds.) {\em Inertial coordinate system on the 
sky} (Dordrecht, Boston, and London: Kluwer), 29-38

\vspace {2 mm}

Howse D 1986 The Greenwich list of observatories: a world 
list of astronomical observatories, instruments and clocks, 
1670-1850 {\em Journal for the History of Astronomy} {\bf 17} 
i-iv, 1-100 (Amendment list no. 1 {\bf 25} 207-218 (1994))

\vspace {2 mm}

Krisciunas K 1988 {\em Astronomical Centers of the World} 
(Cambridge: CUP)

\vspace {2 mm}

M\"{u}ller P 1992 {\em Sternwarten in Bildern: 
Architektur und Geschichte der Sternwarten von den Anf\"{a}ngen bis ca.
1950} (Berlin: Springer Verlag)

\vspace {2 mm}

Needham J and Ling W 1959, {\em Science and civilisation in 
China. Volume 3: Mathematics and the sciences of the heavens and
the Earth} (Cambridge: CUP), pp. 171-461

\vspace {2 mm}

Sayili A 1981 {\em The Observatory in Islam} (New York: Arno 
Press)

\vspace {2 mm}

Thoren V E 1990 {\em The Lord of Uraniborg: a biography of 
Tycho Brahe} (Cambridge: CUP)

\vspace {2 mm}

{\bf Supplementary references:}
\vspace{2 mm}

Deane, Thatcher E. 1989, {\em The Chinese Imperial Astronomical
Bureau: Form and Function of the Ming Dynasty} Qintianjian {\em from
1365 to 1627}, Univ. Washington Dissertation

\vspace {2 mm}

Dreyer, J. L. E. 1890, {\em Tycho Brahe: a picture of scientific
life and work in the sixteenth century}, Edinburgh: Adam and Charles
Black (reprinted 1977 by Peter Smith, Gloucester, Mass.)

\vspace {2 mm}

Hartner, W. 1950, ``The astronomical instruments of Cha-ma-lu-ting,
their identification, and their relations to the instruments of
the observatory of Mar\={a}gha'', {\em Isis} {\bf 41}, 184-194

\vspace {2 mm}

Hermann, Dieter B. 1984, {\em The History of Astronomy from Herschel
to Hertzsprung}, Cambridge Univ. Press

\vspace {2 mm}

Ho Peng Yoke 1997, ``Astronomy in China'', in {\em Encyclopedia of
the History of Science, Technology, and Medicine in Non-Western
Cultures}, Helaine Selin, ed. (Dordrecht: Kluwer), 108-111

\vspace {2 mm}

Huang, Yi-Long 1997, ``Chinese astronomy'' in {\em History of Astronomy:
an encyclopedia}, J. Lankford, ed. (New York and London: Garland), 
pp. 145-149

\vspace {2 mm}

Kaye, G. R. 1973, {\em The Astronomical Observatories of Jai Singh},
Indological Book House, reprint of 1918 edition

\vspace {2 mm}

Krisciunas, K. 1990, ``Pulkovo Observatory's status in 19th century
positional astronomy'', in {\em Inertial Coordinate System on the
Sky}, J. H. Lieske \& V. K. Abalakin, eds. (Dordrecht: Kluwer), pp. 15-24

\vspace {2 mm}

Krisciunas, K. 1992, ``The legacy of Ulugh Beg'', in  {\em Central
Asian Monuments}, Hasan B. Paksoy, ed. (Istanbul: Isis Press), pp. 95-103

\vspace {2 mm}

Krisciunas, K. 1993, ``A more complete analysis of the errors in Ulugh
Beg's star catalogue'', {\em J. Hist. Astron.} {\bf 24}, 269-280

\vspace {2 mm}

Pedersen, O. 1976, ``Some early European observatories'', {\em
Vistas in Astronomy} {\bf 20}, 17-28

\vspace {2 mm}

Sivin, N. 1973, ``Copernicus in China'', in {\em Colloquia Copernicana},
II (Wroclaw: Polskiej Akademii Nauk), pp. 63-114

\vspace { 2mm}

KEVIN KRISCIUNAS

\vspace {1 cm}

\section{Figure captions}

Fig. 1 - Pulkovo Observatory, originally completed in 1839, was
completely destroyed during World War II.  It was rebuilt by 1954.
This is the main building as it appeared in 1989.
(Photo by Kevin Krisciunas)

\vspace {2 mm}

Fig. 2 - Lick Observatory, ca. 1936.  The large white dome houses
the 36-inch refractor, still the second largest in the world.

\vspace {2 mm}

Fig. 3 - Cerro Tololo Inter-American Observatory illuminated by
moonlight.  The largest dome houses the 4-m reflector. (Photo by
Kevin Krisciunas)

\vspace {2 mm}

Fig. 4- The number of optical and infrared telescopes with primary
objectives larger than a given size, ca. 2001.  We include all
telescopes with diameter D greater than or equal to 1.5 meters.
We count the two 8.4-m diameter elements of the Large Binocular Telescope,
the four 8.2-m diameter elements of the Very Large Telescope, and the
two 6.5-m telescopes of the Magellan Project as separate instruments.
Each of the 10-m Keck Telescopes contains 36 hexagonal mirror segments,
while the 11-m Hobby-Eberly Telescope has 91 hexagonal segments. 
(Diagram by Kevin Krisciunas)

\end {document}
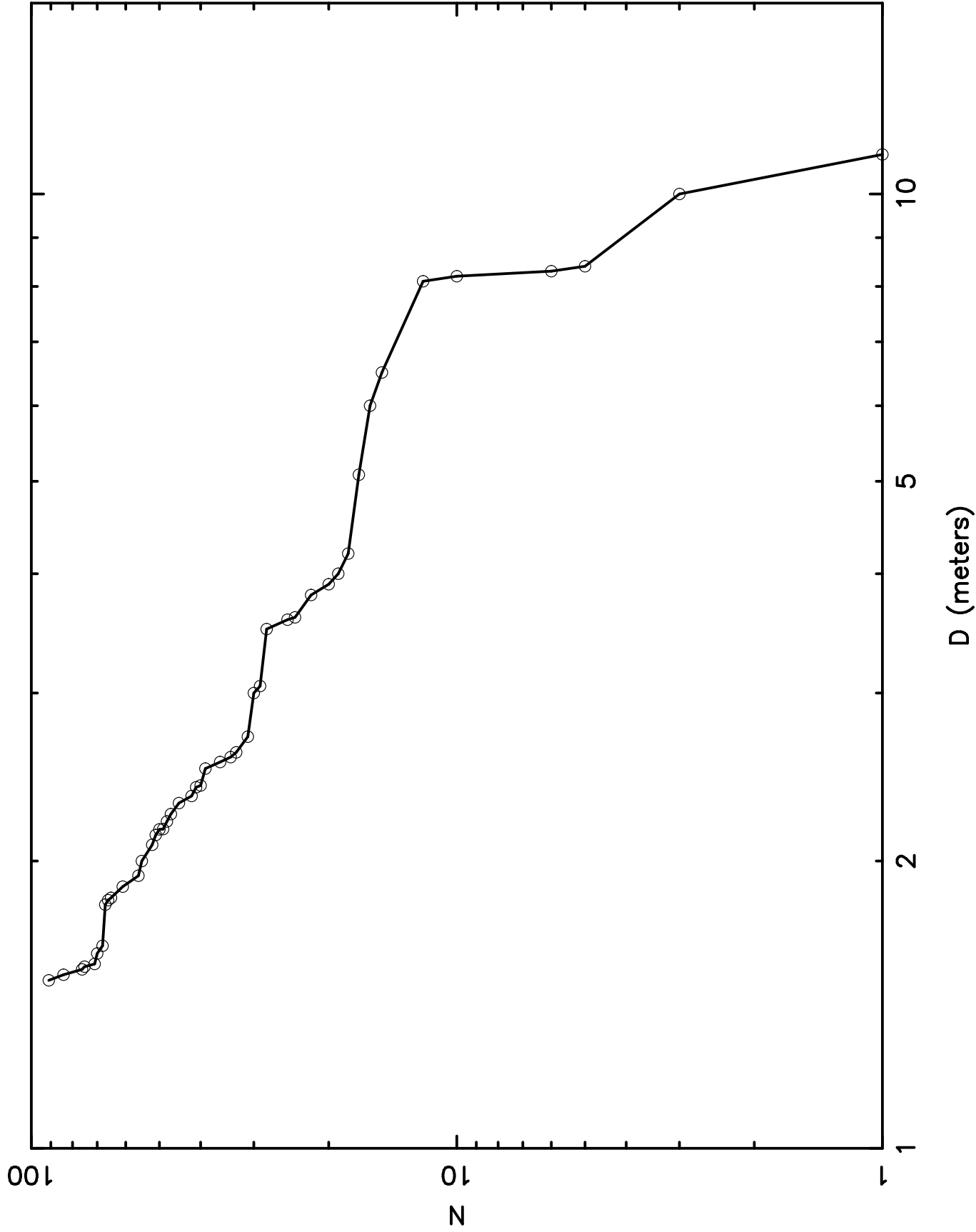
